\documentclass[a4paper,fleqn]{cas-dc}

\usepackage[authoryear,longnamesfirst]{natbib}
\usepackage{hyperref}
\usepackage{subfigure} 
\usepackage{color}
\usepackage{xcolor}
\usepackage{geometry}
\def\tsc#1{\csdef{#1}{\textsc{\lowercase{#1}}\xspace}}
\tsc{WGM}
\tsc{QE}
\tsc{EP}
\tsc{PMS}
\tsc{BEC}
\tsc{DE}

\begin{document}

\let\WriteBookmarks\relax
\def\floatpagepagefraction{1}
\def\textpagefraction{.001}

\shorttitle{An Explainable Ensemble-based IDS for SDVN}

\shortauthors{Ahsan et al.}


\title [mode = title]{An Explainable Ensemble-based Intrusion Detection System for Software-Defined Vehicle Ad-hoc Networks}

\author[1]{Shakil Ibne Ahsan}[orcid=0009-0001-9380-0079]

\cormark[1]

\ead{ahsan026@gmail.com}

\credit{Conceptualisation of this study, Methodology, Software}

\affiliation[1]{organization={University of the West of England}, addressline={Coldharbour Lane}, 
    city={Bristol},
    postcode={BS16 1QY},
    country={UK}}

\affiliation[2]{organization={Intel Labs}, addressline={Intel Corporation}, 
    city={California},
    country={USA}}

\author[1]{Phil Legg}[orcid=0000-0003-3460-5609]
\author[2]{S M Iftekharul Alam}

\credit{Data curation, Writing - Original draft preparation}

\cortext[cor1]{Corresponding author}

\begin{abstract}
Intrusion Detection Systems (IDS) are widely employed to detect and mitigate external network security events. Vehicle ad-hoc Networks (VANETs) continue to evolve, especially with developments related to Connected Autonomous Vehicles (CAVs). In this study, we explore the detection of cyber threats in vehicle networks through ensemble-based machine learning, to strengthen the performance of the learnt model compared to relying on a single model. We propose a model that uses Random Forest and CatBoost as our main ‘investigators’, with Logistic Regression used to then reason on their outputs to make a final decision. To further aid analysis, we use SHAP (SHapley Additive exPlanations) analysis to examine feature importance towards the final decision stage. We use the Vehicular Reference Misbehavior (VeReMi) dataset for our experimentation and observe that our approach improves classification accuracy, and results in fewer misclassifications compared to previous works. Overall, this layered approach to decision-making — combining teamwork among models with an explainable view of why they act as they do — can help to achieve a more reliable and easy-to-understand cyber security solution for smart transportation networks.
\end{abstract}


\maketitle

\section{Introduction}

\begin{figure}
    \centering
    \label{fig:SDVN_diagram}
    \includegraphics[width=0.45\textwidth]{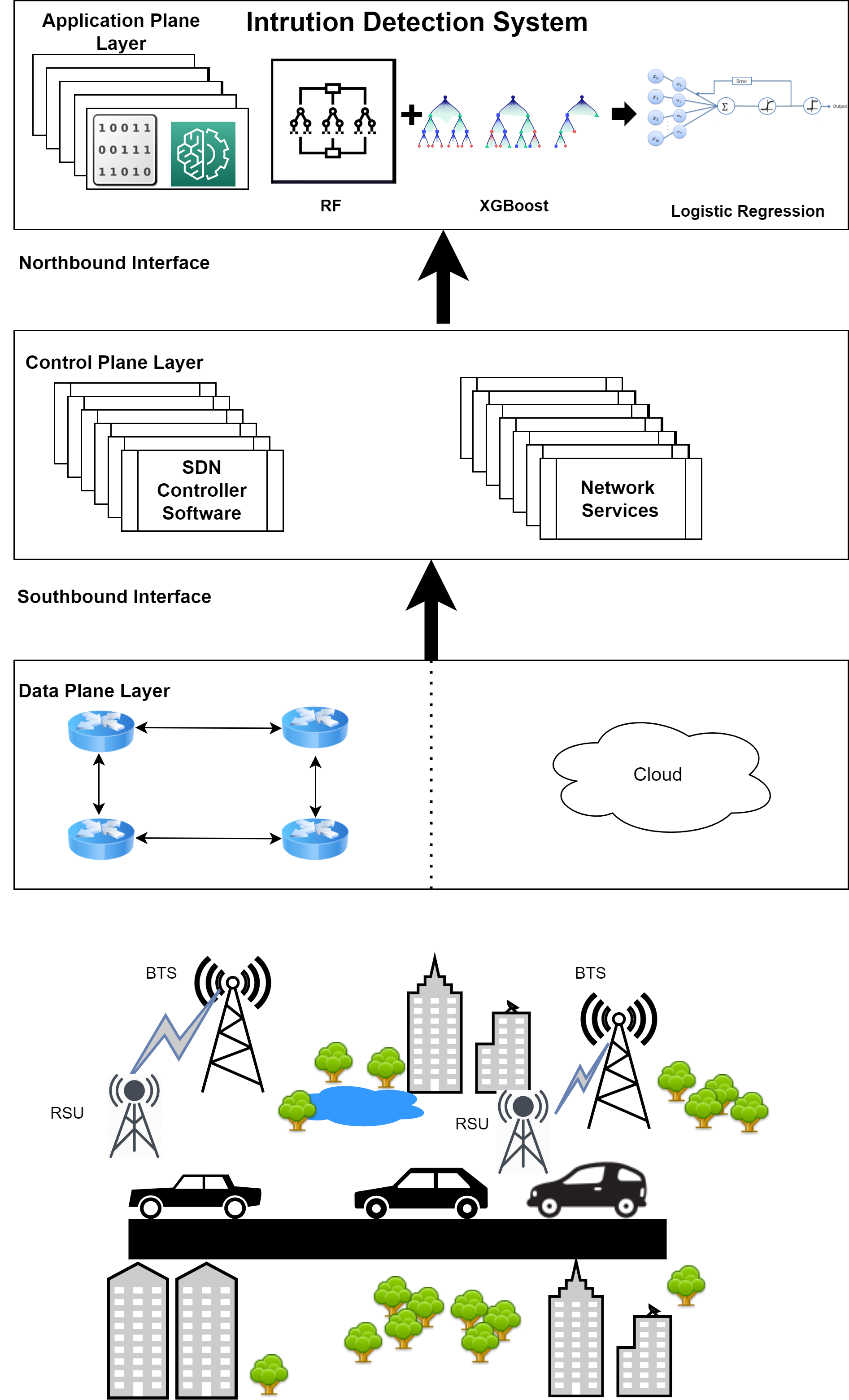}
    \caption{Software-defined VANET}
\end{figure}

As vehicles increasingly connect to their surroundings, Software-Defined Vehicle Ad-hoc Networks (SD-VANETs) are emerging as a vital subset of Mobile Ad-hoc Networks. These networks become particularly relevant in dynamic conditions where the environment shifts due to the constant movement of vehicles. The surge in inter-vehicle communication represents a crucial area of research, standardization, and development, propelled by the growing incorporation of wireless communication and computing technologies in automobiles. This integration forms the backbone of SD-VANETs (Figure \ref{fig:SDVN_diagram}), which facilitate wireless connections between both stationary and mobile nodes, enabling seamless communication that enhances road safety and efficiency.

However, this advancement has a serious downside: it opens up a new battleground for cyber threats. Automotive cybersecurity is not just a technical challenge—it is a matter of public safety, privacy, and the stability of entire transportation systems. While traditional Intrusion Detection Systems (IDS) can catch many threats, they often fall short where it matters most: explaining why certain threats are flagged or missed, especially in fast-paced, real-time environments.

Interpretability in IDS models is essential. First, VANETs are dynamic, with high-speed vehicles and constantly shifting network conditions. Security systems for these environments need to be more than accurate—they need to be transparent, so operators can trust and quickly act on what the system detects. Second, a model that can explain itself allows designers to see exactly where it might go wrong, giving them a way to strengthen it and cut down on false positives. In high-traffic vehicle networks, even minor errors can lead to major problems, so interpretability is non-negotiable.

This study tackles this critical gap by introducing an innovative IDS framework that combines Explainable Artificial Intelligence (XAI) with SHAP (SHapley Additive exPlanations) analysis. This approach goes beyond simply detecting threats; it shows why certain patterns are detected, providing clarity and reliability in the complex landscape of VANET security. Our focus on interpretability alongside performance sets a new standard in cybersecurity for intelligent transportation, meeting the real-world demands of this high-stakes field.

By referring to the literature review below, several essential study gaps in ML-based IDS research have been noted. Still, there is enough opportunity to improve the accuracy of the intrusion detection model. Moreover, IDS users still struggle to detect known unencrypted attacks because IDS has not yet achieved 100\% accuracy. One significant issue with IDS is that they regularly alert the auditor to false positives. However, an IDS can be optimised to reduce the number of false positives and false negatives, which is a deficiency of publicly available VANET databases with attacks. No fast, inexpensive, and effective in-built privacy-preserving ML-based IDS is still visible. VANET is a type of wireless network where both stationary and mobile nodes can connect. Such a system has frequent disconnections, a dynamic topology, and significant node mobility. Each attribute may result in a security flaw that hazards the system. Due to the high mobility of nodes, VANET topology is dynamic and unpredictable; this and other circumstances, such as severe weather conditions, may cause VANET to have several disconnections throughout the system.

\subsection{Contributions} 

The contributions of this paper are as follows: 

\begin{enumerate}
    \item This research proposes a complimentary ensemble approach that integrates Random Forest, CatBoost, and Logistic Regression, providing a suitable trade-off between accuracy, explainability, and runtime requirement for intrusion detection within SDN-based VANETs. 
    \item We demonstrate our approach against the Vehicular Reference Misbehaviour (VeReMi) dataset to illustrate the recognition of different attack classes within VANETs. Furthermore, the integration of XAI using SHAP analysis, provides an explanability of how individual models and their collective ensemble can inform decision-making, and in particular, for understanding the model reasoning behind remaining misclassification. We chose Random Forest and CatBoost because they are particularly good at handling high-dimensional and categorical data. This capability is crucial for managing the complexity and variability of data found in VANETs.
\end{enumerate}

\section{Background}\label{sec2}

VANET, a type of wireless network connecting both stationary and mobile nodes, is characterised by frequent disconnections, a dynamic topology, and high mobility, which can lead to potential security flaws. Compared to Mobile Ad Hoc Networks (MANET), VANETs exhibit considerably greater mobility and more stringent topology restrictions. Vehicular Ad Hoc Networks (VANETs) enable applications such as real-time traffic monitoring, blind crossing prevention, dynamic route planning, collision prevention, and internet access for vehicles. According to the World Health Organization (WHO), 1.3 million people die in road crashes annually, with 20 to 50 million suffering non-fatal injuries (\cite{World2021}).

The new era of automobiles and road traffic systems will depend on wireless networks and machine learning. As stated previously, many lives have been lost on highways. Consequently, when road traffic increases, road safety must also grow. We need a mechanism by which vehicles can be made sufficiently intelligent to manage their process for road safety. This notion constituted the foundation for VANETs. SDN-based VANETs follow the below architecture (\cite{Ku2014}). 

\subsection{Communications And Standards}

The USDOT (The United States Department of Transportation) has designated Dedicated Short-Range Communication as the  preferred inter-vehicle communication medium. The communication infrastructure for linked automobiles is provided via WAVE protocols. IEEE 802.11p protocol and IEEE 1609.x protocols define WAVE. Their activities conform to the internet protocol stack. Physical and data connection layers are supported by IEEE 802.11p. The tiers of operation for IEEE 1609.x are network, transport, and application. This protocol focuses on the physical and Media Access Control (MAC) levels to provide low latency and good dependability across short radio communication lines. It functions at both the physical and MAC levels. It offers a direct wireless interface and mapping between MAC and physical data. In addition, it provides media access rules and aids in multi-channel operations at the MAC layer (\cite{Chekkouri2015}). 

\subsubsection{IEEE 802.11p} 

IEEE 802.11p enables security via authentication and encryption of private communications. All cars are expected to be registered by a Motor Vehicle registration unit. This would assign MAC IDs to automobiles. Alternatively, Vehicle Identification Numbers (VINs) can serve as MAC addresses for automobiles. Vehicles will obtain IP addresses upon network connection. Vehicle identification will be based on their MAC addresses, IP addresses, or publicly provided IDs (\cite{Zheng2016}).

\subsubsection{IEEE 1609.4p} 

As highlighted in (\cite{Chen2009}), IEEE 1609.4p support for multi-channel operations is achieved by time division so radios can switch between different channel intervals. Control Channel (CCH) and Service Channel (SCH) are the two channels of this IEEE standard. Continuous safety messages are sent via CCH, whilst other applications are routed via SCH. IEEE 1609.3p defines the network and transport layer services. Here, IP addressing and routing occur.
IEEE 1609.2p is the protocol that specifies secure message exchanges and encryption. IEEE 1609.1p is a protocol specifying alert and probe vehicle message forms. Currently, a modification to IEEE 1609.x is in the works. The following enhancements have been made. Draft IEEE 1609.5 to handle vehicle-to-vehicle and vehicle-to-infrastructure connectivity. Draft IEEE 1609.6 to provide remote management between Onboard Units and Road-Side-Units. IEEE Standard 1609.11 for safe electronic transactions.IEEE Standard 1609.12 defines the allocation of value identifiers in WAVE. 

\subsection{SDN strategies to address VANET security}

SDN can manage network-wide transmissions without configuring devices separately, providing a holistic view and centralised control. These features enhance the security of Vehicular Ad Hoc Networks (VANETs) by defending against common vulnerabilities. \cite{AKHUNZADA2016}.
Nevertheless, the network will be most susceptible to security concerns if a rogue node compromises or impersonates the controller. An attacker may issue commands to network devices to undertake malicious actions through the hacked controller. Through programmability, attackers can build false and inaccurate flow rules to interrupt communication and manipulate devices and software. These harmful operations are capable of causing network resource unavailability, accidental occurrences, traffic congestion, and communication delays. The following section discusses various risks associated with SDN-based VANET. These two conceptions are at odds with whether SDN is a security benefit for VANET or a security risk for VANET. 

\subsection{Intrusion Detection System (IDS)}
The intrusion detection system is a network-based system that detects any harmful activity on the target network. The software architecture operation depends on observing the network and noting any attempts or incidences of network infiltrations into the system as the necessary actions are done in time to prevent the systems from getting harmed. When an IDS identifies malicious behaviour in a network or node, it also blocks the malicious node's access to the network to avoid additional network damage. Intrusion Detection System designs can be centralised or decentralised. A centralised intrusion detection system collects and analyses data in a centralised system, whereas a distributed intrusion detection system collects and analyses data on several dispersed hosts. A centralised IDS achieves effective intrusion detection for a single system. The sites for data analysis remain constant regardless of the number of observed hosts. In situations involving numerous systems, IDS employs a decentralised design. In decentralised intrusion detection systems, numerous data security analyses are performed at several network locations. The number of places where data analysis is conducted is proportional to the number of observed host computers. This is the practical system architecture achieved by evaluating the network and detecting any intrusion attempts or malicious activities that made the network vulnerable so that appropriate measures could be taken promptly to prevent such activities from causing damage. There are five distinct categories of VANET attacks: (1) authenticity, (2) confidentiality, (3) availability, (4) integrity, (5) accountability, and (6) privacy. The Intrusion Detection System counteracts harmful actions by blocking a questionable IP address, port, or person from accessing and gaining access to the network. Depending on the goal of intrusion detection systems, they are separated into three types. 
\newline 

A network-based intrusion detection system (NIDS) operates at the network level to identify any rogue network node or incursion attempt in the network. NIDS are strategically positioned within the network to manage traffic to and from all equipment. It analyses passing traffic on the whole subnet and compares it to a library of known assaults. The administrator can be notified once an attack or suspicious behaviour is detected.
Host-based Intrusion Detection Systems (HIDS) function locally on a particular node to detect intrusions. HIDS operate on particular network hosts or devices. A HIDS analyses only the inbound and outbound packets from the device and alerts the user or admin if any unusual behaviour is noticed. It compares the current snapshot of system files to the past snapshot. If vital system files are updated or deleted, the administrator is notified to examine them.
Wireless Intrusion Detection Systems (WIDS) operate wireless networks with functionality comparable to NIDS. Based on their detecting strategies, IDS is categorised under Misuse Detection and Anomaly Detection and, based on the behaviour of IDS.

\section{Related Works}\label{sec3}

As a part of the research, different articles have been reviewed to explain how earlier researchers have ended work on an Intrusion Detection System for cyber-attack detection in SDN-based VANET. This Literature review study employs the methods stated by Preferred Reporting Items for Systematic Reviews and Meta-Analyses (PRISMA). The methodology covers eight main steps: User needs assessment, deﬁne the Research Questions, Search String, Selection of Source, Select Relevant Paper, Screening Findings, Extract Data, and ﬁnally, synthesise the collected data (\cite{Kitchenham2007-dg}).

IDS and Malware detection systems can still not detect unknown, Encrypted, and Inference attacks with 100\% accuracy. According to Kaspersky, most Malware detection products do not even reach the required 90\% security threshold. A significant player in the IDS market, CISCO, claimed that Rapid detection—those made in less than five minutes—and long-term learning capacities displayed excellent performance in their IDS. Cisco Encrypted Traffic Analytics, or IDS, discovered nearly two-thirds of all known malicious flows in less than five minutes. Moreover, Cisco declared this IDS demonstrated better identification results even with low volumes of 0 to 20 flows, whereas 2000 plus flows enabled 100\% detection for known attacks. Nevertheless, this solution has a limitation with a small attack size and taking three hours to gain 100\% accuracy. However, the Cisco Adaptive Security Device Manager (ASDM) to manage Cisco Secure Firewall ASA (with IDS/IPS) has been impacted by Common Vulnerabilities and Exposures CVE-2022-20829, CVE-2021-1585, and CVE-2022-20828. Researcher Jake Baines, who disclosed them to Cisco in February and March of 2022, added that many users were undoubtedly not updating their Cisco firewalls properly (\cite{Cisco2022} and \cite{Scroxton2022}).

\subsection{Intrusion detection system challenges in Software-defined vehicular network (SDVN)}

There are several difficulties in creating an adaptable and versatile IDS using ML in SDN-based VANET networks due to the need for effective data classification in academic research. To have a consistent and accurate evaluation of intrusion detection systems, it is essential to construct robust datasets. For testing and evaluation of intrusion detection, the following datasets are already available: 

\subsubsection{VeReMi} 
According to \cite{Van_der_Heijden2018-ah}, the VeReMi dataset was created to test the effectiveness of VANET misbehaviour detection systems, based on the open-source VEINS vehicular simulation framework. This dataset can be efficiently employed to test the performance of misbehaviour detection approaches implemented on a city-wide scale while utilising fewer computing resources than standard (frequent) VEINS simulations. The dataset comprises On-Board Unit (OBU) message logs, including timestamps, sender IDs, positions, speeds, and reception times. VeReMi also includes a verification file, which provides GUI parameters on the validity of the data of each message upon which the verification confirms whether it is a ``normal'' or ``attack'' case. The dataset spreads across three attacker densities, five distinctive attack types, and three levels of density; it also includes various scenarios at different times of the day, including how traffic varies across the time of the day with the vehicle density range from 35 vehicles in low-dense areas to 519 vehicles during maximum congestion; and diverse types of attacks where the attacks in this paper are attacks against the position expressed in many ways from the position remaining still to noising out the position. While versatile, the VeReMi dataset is restricted in terms of both position-falsifying attacks and all types of network threats and advanced malicious strategies. Being a synthesised dataset, it lacks real-life cases but can be developed further by researchers.

\begin{table}[tbp]
\caption{VeReMi Dataset attacks and size.}
\label{table:veremi}
\begin{tabular}{@{}ll@{}}           \\ \hline
Attacks                                & Size   \\ \hline
BENIGN                                 & 437429 \\
Attack type 16 (Eventual stop Attack)  & 56595  \\
Attack type 4 (Random Attack)          & 30510  \\
Attack type 2 (Constant Offset Attack) & 30473  \\
Attack type 1 (Constant Attack)        & 30473  \\
Attack type 8 (Random Offset Attack)   & 29460  \\
\hline
\end{tabular}
\end{table}

The dataset is publicly available for reproducibility; thus, the research results made from the dataset are reliable and verifiable. VeReMi has 562,467 entries, out of which 437,429 entries are benign and 125,038 are attacking entries. The dataset comprehensively covers a city-level context of various types of roads and traffic behaviours, thus making it a reliable framework for benchmarking and testing the performance of misbehaviour detection approaches in VANETs (\cite{Van_der_Heijden2018-ah}). Table \ref{table:veremi} shows the data classifications and the number of instances that occur within the dataset.

\subsubsection{Attack detection techniques}

The work by \cite{Singh2018} discusses the ML-based tactics to detect attacks in SDN-based vehicular networks based on machine learning techniques. The research has a particular emphasis on vehicle-to-infrastructure (V2I) links. It analyses the current environment to establish the ML strategy that will most successfully detect threats posed by distributed denial of service (DDoS) attacks (Short-lived Transmission Control Protocol Flooding, User Datagram Protocol Flooding). To accomplish this, it collects seven different elements of actual data from TCP (Transmission Control Protocol) and UDP (User Datagram Protocol) flows both when DDoS attacks them and when they are not. The network is educated by employing several different supervised learning algorithms depending on the data acquired, and it is discovered that the gradient boost classifier produces the most satisfactory outcomes. In addition, the random forest, decision tree, and linear support vector machine (linear-SVM) approach all produce results that are close to being the best. On the other hand, the limited training data could be to blame for the weak performance of Neural Networks (NN). Accurate, measurable data was not found in the write-up or the chart.

According to \cite{Aneja2018}, they presents an AI-based system to identify a Route Request (RREQ) flooding attack. Researchers used a genetic algorithm (GA) to optimise feature subset selection and an Artificial Neural Network (ANN) for classification. The model scored 99\% accuracy from  Network Simulator 2 (NS2) 1000 entries simulated dataset. 

In their study, \cite{Yu2018} analyse DDoS assaults in an SDN-based vehicular network by employing TCP flood, UDP flood, or ICMP (internet control message protocol) flood. In this research paper, the authors present a scheme for detecting Distributed Denial-of-Service (DDoS) attacks in SDVN atmospheres using three separate models. A ``trigger detection model'' for inbound packets and another method is the ``flow table feature-based detection model''. Both methods use the OpenFlow protocol features and the detection of a threat model based on support vector machine classification. All of these models are used in conjunction with one another. They synthesised and simulated network traffic to accommodate various DDoS attacks via Scapy and hping3, and the findings suggested an efficiency of over 97\%. In the article, ML-Based Intrusion Detection Method for VANETs.

In their research, \cite{Zeng2018} developed an ML-based Senior2Local IDS with propagation to identify potential threats in VANETs both Globally and Locally. They used a fuzzification ANN technique to recognise malicious attacks globally and protect Cluster Heads in Roadside Units (RSUs). After that, a game theory trust system or RSU is developed separately. They employed a simple and straightforward Support Vector Machine (SVM) to detect attacks locally. The findings demonstrated that the method was more reliable and resilient than earlier ML-based tactics.

As outlined by \cite{Gad2021}, the researchers conducted the experiment on a ToN-IoT dataset and tested various ML approaches in binary and multiclass classification problems in VANET to detect intrusion. On the other hand, LR is used for solving binary classification problems, while multiclass issues are addressed with a one-vs-rest approach.

\cite{LIU2014} suggests employing data mining methods in VANETs to identify existing assaults and uncover previously unidentified attacks. The proposed approach makes three primary contributions. First, it develops a decentralised vehicle network to enable scalable communication. Second, it uses two data mining models to demonstrate the viability of an IDS in VANETs. Lastly, it discovers novel patterns of previously unidentified threats.

The suggested solution uses a cell grid to organise the previously mentioned network. Every cell has a transmission tower, enabling users to communicate with others and the internet. Each will use data mining models and rules to identify new assaults. Therefore, the IDS can develop new rules that may be communicated for each subnetwork. The trucks and the tower cell collect the data sent throughout the network. The authors use Naive Bayes and Logistic Regression classifiers to analyse the obtained data. Finally, the authors use simulation to evaluate their strategy's performance over a network.

First, the Simulation of Urban Mobility (SUMO) simulator produces a file that tracks movement. Then, to mimic a wireless network, this information is given into ns-2. This scenario includes 150 cars that turn randomly at junctions, drive at the posted speed limit (which ranges from 5 to 20 meters per second), and stop at traffic lights that are randomly positioned. Five automobiles were outfitted with Linux to validate the IDS and made to run various network applications. After that, nine months were spent using TCPdump to generate the dataset. There were four different kinds of attacks, with 39 different types documented. After that, WEKA is used to categorise them. Finally, the metrics recall, F-measure, and Matthews's correlation coefficient were utilised to assess the models.

\subsubsection{Methods yet to reach optimum accuracy}

As reported by \cite{Türkoğlu2022}, with the help of feature selection and hyperparameter optimisation, the DT classifier attained an accuracy score of 99.35\%, which was the best possible result. The excellent results achieved with the methodology developed proved the considerable significance of feature selection and optimisation techniques. These findings were demonstrated by the outstanding results obtained with the method.

In the findings of \cite{Kim2017}, a collaborative security threat detection method is demonstrated within the software-defined vehicular cloud architecture. This method highlights the importance of collaborative approaches in enhancing security within vehicular networks. A traditional vehicular environment has few exceptional features, such as high flexibility, dynamic networking, and rapid connectivity. His proposed model said a Speed Dependent Volume Controller (SDVC) had collected the flow information from the vehicular cloud (VC) to train multiclass SVM. In the next step, the SVM classifier identifies the bad actors among the network's nodes. However, performance is still under 90\% for Accuracy score, Precision, and Recall. 

However, \cite{Misra2011} offers an LA-based system for IDS in VANET. The suggested methodology protects users' anonymity by giving each car a unique, changing ID. The proposed approach involves developing a VANET administration system. There is a base station at the beginning and end of each route, and all cars have transmitters to stay in touch with the base stations. To redirect traffic, each attacker will generate malicious packets. The budgets of the attackers and the system are similar. The more the attacker's resources, the greater the number of packets it can create. All that is needed to identify the attackers is their dynamically assigned identifiers. If unique IDs are assigned to vehicles, and many vehicles have the same ID, then at least one of the vehicles is likely malevolent. Under attack, grid budgets may be attributed, and sampling rates can be recalculated with the help of learning automata. Simulations run on their simulator are used to evaluate models. The success rate for detecting four attackers ranges from around 44\% to 77\%. In the suggested setup, the IDS is located in each base station and has complete sight of overall network traffic. 

\cite{Tian2010} proposes a Bus Ad-Hoc Network (BUSNet) based IDS tailored to protect VANETs from malicious activity. This hierarchical anomaly detection technique is built on the virtual setup of the mobile bus agents. In this configuration, the buses serve as the cluster heads, collecting and relaying data packets broadcast by passing vehicles through accessing points installed along the roadways. After that, an NN-based classification system takes that data and uses it to spot DDoS assaults. The proposed method is tested in an artificial network constructed with the ns-2 network simulator. The researchers do not disclose the method or traffic simulator utilised to generate the simulation results. Fifty unique vehicles are simulated, each communicating using a constant bit rate (CBR) and 512-byte packet size. During each attack, two vehicles work together to send out updates at small durations of 0.01 seconds at four different checkpoints over 10 seconds. Both vehicles are used in every attack. The authors establish their benchmark by picking a value below which they treat everything as an assault. This number might range from 0.05 to 0.7, which helps to explain the discrepancy in these results. A threshold value of 0.2 is recommended as the best option. Overall, accuracy has not achieved the milestone to protect VANET.

According to findings of  \cite{Mejri2014}, the proposed solution is a detection system that employs a statistical method, specifically linear regression, along with a watchdog mechanism to passively identify greedy behaviour at the MAC layer. A watchdog monitors the degree to which active nodes' access periods correlate as part of the suggested approach. The network is under assault if the correlation coefficient is close to 1 but not precisely 1, and the linear regression straight-line slope is also not close to 1. This can happen when the correlation coefficient is near one but not quite one or when the slope is very close to 1 but not quite 1. The developed program does this by monitoring three metrics: the interval between transmissions, the total time spent transferring data, and the number of connect attempts made by a node. Subsequently, the performance of the solutions is analyzed in real-time using a trace file generated by SUMO. The SUMO scenario is constructed based on a city map that includes labels and traffic signals. To validate the reliability of the linear regression method, researchers initially simulate the network with nodes exhibiting correct behaviour. After that, four greedy automobiles are produced by injecting greedy nodes one at a time. To detect greedy behaviour, this approach takes 1.3 seconds for 1 node, 1.9 seconds for 2 nodes, 3.1 seconds for 3 nodes, and 7.9 seconds for 4 nodes. \cite{Alheeti2015} The solution proposed in the paper is an ANN-based ill-use and anomaly intrusion detection system that can identify DoS assaults. The authors designed the solution by first developing a mobility scenario within SUMO. For ns-2 to read the files SUMO created, a conversion was performed using Mobility Vehicle (MOVE). In the end, ns-2 was utilised to imitate the vehicles' communication ability. When considering the mobility and traffic situation, the authors used the urban mobility of Manhattan as their inspiration. The simulation takes two minutes and fifty seconds, and it involves cars and six Road roadside units (RSUs). A CBR (Constant Bit Rate) traffic that conveys User Datagram Protocol (UDP) packets is being performed by the vehicles. Only one of the automobiles intentionally tries to cause harm. It will discard the packets rather than send them to the destination. The solution built was accurate 85.02\% of the time when classifying normal behaviours and 98.45\% when classifying deviant behaviours. 

The authors of \cite{Gruebler2015} advocate using a Proportional Overlapping Scores (POS) approach to reduce the number of features taken from the trace file. These are fed into an artificial neural network (ANN) to teach it to classify data. After that, the ns-2 trace files that were created were employed in the POS algorithms so that the appropriate characteristics could be extracted. Next, Fuzzification is applied to the dataset's data to prevent classification difficulties. Finally, the dataset is classified with the help of a Feed-Forward Neural Network (FFNN) using the IDS created for this solution. Sixty thousand records from the dataset were used, with half designated for training, another quarter for testing, and the remaining quarter for validation (25\%). This dataset includes both benign and harmful activity in equal measure. The malicious behaviour vehicle was designed as a Black Hole attack, in which the malicious vehicle would not forward any received packets but dump them instead. Ultimately, the intrusion detection system was evaluated using anomaly and misuse detection. Regarding identifying misuse, it acquired a classification of 99.89 \% for normal conduct and 99.80\% for deviant behaviour. Regarding detecting anomalies, the findings for typical and aberrant behaviours came in at 99.87 and 99.72\%, respectively. 

In their paper, \cite{Sedjelmaci2015} introduce a cluster-based intrusion detection system (IDS) to defend a network against selective forwarding, black holes, wormholes, packet repetition, resource depletion, and Sybil attacks. The technique that has been presented uses several different detection managers that operate on three different levels: cluster associate, cluster head, and RSU. The cars' speeds determine how they should be organised into clusters. Cluster connection and the assurance of data safety are the criteria utilised in selecting the cluster heads. When social behaviour is introduced, the connectivity inside a cluster is improved. The IDS architecture comprises two primary detection schemes and a decision scheme. Local Intrusion Detection System and Global Intrusion Detection System are the detection systems run at cluster members and heads. The name of the decision-making scheme is Global Decision System, which is hosted on the RSUs. Because of this, the system can monitor several organisations and identify threats on multiple levels. In addition, each level can perform a unique set of algorithms and detecting approaches, which evaluate a unique set of characteristics. The Global Decision System will provide the collective standing of each car, which the cluster head will then send. The Worldwide Decision Scheme will then calculate the trust level of every vehicle. To evaluate the proposed system, ns-3 was used as the network simulator, while SUMO was used to mimic the vehicle's movement. The IDS was evaluated concerning the detection rate, the false positive rate, and the detection time. The various assaults were put through their paces using a range of vehicle counts during testing. The percentage of hostile cars always equals 45\% of the total. The provided findings range from 92\% to 100\%, although this depends on the total number of vehicles and the type of attack conducted. 

In their research, \cite{Alheeti2016} propose an FFNN and SVM-based intrusion detection system is suggested in this paper. A methodical reaction is also meant to protect automobiles if harmful conduct is identified. A dataset composed of SUMO and ns-2 is utilised to train the IDS, and the features extracted from the trace file are used as training data. The POS algorithm has reduced these characteristics from 21 to 15.~The collection consists of 30,000 entries that differentiate between appropriate and inappropriate behaviours. Fuzzification is performed on this dataset before it is used for classification. The writers separate the dataset into three categories: validation, test, and training. Generating grey-hole attacks involves picking malicious vehicles that drop packets at arbitrary intervals throughout the attack. The Ad Hoc On-Demand Distance Vector (AODV) protocol needs to be modified so that rushing assault can be carried out successfully. In SVM, the accuracy values achieved through simulation were 99.93\% for normal behaviour and 99.64\% for aberrant conduct. When FFNN was used, the findings showed an accuracy of 99.82\% for normal behaviour and 98.86\% accuracy for deviant conduct, respectively.

\begin{table*}[tbp]
\fontsize{5}{5}\selectfont
\caption{Review of different ML-based IDS models in VANET and SDVN.}
\label{table:review}
\begin{tabular}{cccccc}
\hline
Reference &
  ML Models &
  Dataset &
  Attack &
  Featuring Method &
  Top Result\\ \hline
\begin{tabular}[c]{@{}c@{}}\cite{Singh2018}\end{tabular} &
  \begin{tabular}[c]{@{}c@{}}GB, RF, DT, \\ Linear SVM, \\ LR, NN,\\ Neural Net.\end{tabular} &
  \begin{tabular}[c]{@{}c@{}}Generated \\ (14456)\end{tabular} &
  \begin{tabular}[c]{@{}c@{}}Short-lived \\ TCP Flood, \\ UDP Flood.\end{tabular} &
  - &
  \begin{tabular}[c]{@{}c@{}}Not measurable \\ from the chart\end{tabular} \\
\begin{tabular}[c]{@{}c@{}}\cite{Aneja2018}\end{tabular} &
  ANN &
  \begin{tabular}[c]{@{}c@{}}NS2 Trace \\ (1000)\end{tabular} &
  RREQ Flood &
  GA based &
  \begin{tabular}[c]{@{}c@{}}Accuracy: 95\%, \\ Precision: 97\%, \\ F-Score: 98\%, \\ FPR: 3\%\end{tabular} \\
  \\
\begin{tabular}[c]{@{}c@{}}\cite{Yu2018}\end{tabular} &
  SVM &
  \begin{tabular}[c]{@{}c@{}}DARPA 2000, \\ CAIDA \\ DDoS 2007\end{tabular} &
  \begin{tabular}[c]{@{}c@{}}TCP ﬂood, \\ UDP ﬂood, \\ ICMP Flood,\\ Etc.\end{tabular} &
  \begin{tabular}[c]{@{}c@{}}Flow-based \\ features \\ and statistical\\ based entropy\end{tabular} &
  \begin{tabular}[c]{@{}c@{}}TCP Traffic \\ Pre-Selection: \\ DR 98.26 \%, \\ FR 0.52\% \\ After-Selection: \\ DR 98.56\% \\ FR 0.32\%\end{tabular} \\
  \\
\begin{tabular}[c]{@{}c@{}}\cite{Zeng2019}\end{tabular} &
  \begin{tabular}[c]{@{}c@{}}DeepVCM, \\ KNN, DT, \\ 1D-CNN, LSTM\end{tabular} &
  \begin{tabular}[c]{@{}c@{}}ISCX2012, \\ NS3 Trace \\ (28236)\end{tabular} &
  \begin{tabular}[c]{@{}c@{}}Brute Force,\\ SSH DDoS,\\ HttpDoS\end{tabular} &
  RAW data &
  \begin{tabular}[c]{@{}c@{}}NS3 Dataset/\\ DeepVCM/DoS \\ Precision 98.5\% \\ Recall 98.3\% \\ F1-Score 98.4\%\end{tabular} \\
  \\
\begin{tabular}[c]{@{}c@{}}\cite{Zeng2018}\end{tabular} &
  ANN &
  Generate &
  - &
  \begin{tabular}[c]{@{}c@{}}CNN \\ algorithm \\ applied to \\ Extract features.\end{tabular} &
  \begin{tabular}[c]{@{}c@{}}Accuracy: \\ SMV-CASE 98.7\%, \\ CEAP 98.9\%, \\ Senior2Local 98.37\%\end{tabular} \\
  \\
\begin{tabular}[c]{@{}c@{}}\cite{Gad2021}\end{tabular} &
  \begin{tabular}[c]{@{}c@{}}LR, NB, \\ DT, SVM, \\ kNN, RF, \\ AdaBoost, \\ XGBoost\end{tabular} &
  ToN-IoT &
  \begin{tabular}[c]{@{}c@{}}DoS\\ DDoS\\ etc.\end{tabular} &
  \begin{tabular}[c]{@{}c@{}}Features \\ Selection: \\ a. Apply Chi² \\ b. Apply \\ SMOTE \\ c. Apply above \\ both\end{tabular} &
  \begin{tabular}[c]{@{}c@{}}With All Features\\  XGBoost: \\ Accuracy 99.1\%, \\ Precision 98.4\%, \\ Recall 99.1\%, \\ F1-Score 98.7\%\end{tabular} \\
  \\
\begin{tabular}[c]{@{}c@{}}\cite{Türkoğlu2022}\end{tabular} &
  \begin{tabular}[c]{@{}c@{}}kNN, SVM\\ DT\end{tabular} &
  Generated &
  DDoS &
  \begin{tabular}[c]{@{}c@{}}MRMR \\ filter \\ method\end{tabular} &
  \begin{tabular}[c]{@{}c@{}}DT Algorithm with \\ 25 Features: \\ Accuracy 99.35\%, \\ Sensitivity 99.22\%, \\ Specificity 99.80\%, \\ F1-Score 99.21\%\end{tabular} \\
  \\
\begin{tabular}[c]{@{}c@{}}\cite{Kim2017}\end{tabular} &
  SVM &
  KDD CUP 1999 &
  \begin{tabular}[c]{@{}c@{}}DoS, Probing, \\ U2R, R2L \\ Etc.\end{tabular} &
  - &
  \begin{tabular}[c]{@{}c@{}}Under Diff. Attack \\ (Approx. from \\ Graph when $\alpha$=0): \\ Accuracy \\ SVM-VC 68.00\%, \\ SVM Nearest \\ Neighbour 57.00\%, \\ SVM Individual 39.50\%\end{tabular} \\
  \\
\begin{tabular}[c]{@{}c@{}}\cite{LIU2014}\end{tabular} &
  NB, LR &
  TCPdump &
  \begin{tabular}[c]{@{}c@{}}DoS,   R2L, \\ U2R, Probing\end{tabular} &
  - &
  \begin{tabular}[c]{@{}c@{}}Not measurable \\ from Chart. \\ LR showed the \\ best result.\end{tabular} \\
\begin{tabular}[c]{@{}c@{}}\cite{Alheeti2015}\end{tabular} &
  ANN &
  NS2 Trace ﬁle &
  DoS &
  - &
  Accuracy 85.02\% \\
  \\
\begin{tabular}[c]{@{}c@{}}\cite{Gruebler2015}\end{tabular} &
  ANN &
  \begin{tabular}[c]{@{}c@{}}NS2 Trace ﬁle and\\ Animator (32000)\end{tabular} &
  Black Hole &
  - &
  Accuracy 85.02\% \\
  \\
\begin{tabular}[c]{@{}c@{}}\cite{Alheeti2016}\end{tabular} &
  \begin{tabular}[c]{@{}c@{}}FFNN,\\ SVM\end{tabular} &
  NS2 Trace ﬁle &
  \begin{tabular}[c]{@{}c@{}}Grey Hole,\\ Rushing\end{tabular} &
  - &
  \begin{tabular}[c]{@{}c@{}}Accuracy:\\ SVM-Abnormal \\ 99.80\%\end{tabular} \\
\begin{tabular}[c]{@{}c@{}}\cite{Sharma2018}\end{tabular} &
  SVM &
  NS2 Trace ﬁle &
  \begin{tabular}[c]{@{}c@{}}Wormhole, \\ Selective Forwarding,\\ Packet Drop\end{tabular} &
  \begin{tabular}[c]{@{}c@{}}Hybrid \\ Fuzzy \\ Multi-\\ Criteria \\ Feature\\ Selection\end{tabular} &
  \begin{tabular}[c]{@{}c@{}}Performance \\ evaluated \\ by DR, FPR, DT\end{tabular} \\
  
\begin{tabular}[c]{@{}c@{}}\cite{Kumar2015}\end{tabular} &
  \begin{tabular}[c]{@{}c@{}}Learning \\ automata\end{tabular} &
  Generated &
  Flooding, Blackhole &
  - &
  - \\
\begin{tabular}[c]{@{}c@{}}\cite{Shu2021}\end{tabular} &
  \begin{tabular}[c]{@{}c@{}}MLP (ANN), \\ BiGAN\end{tabular} &
  \begin{tabular}[c]{@{}c@{}}KDD99, \\ NSL-KDD\end{tabular} &
  \begin{tabular}[c]{@{}c@{}}DoS, U2R,\\ R2L, Probing etc.\end{tabular} &
  - &
  \begin{tabular}[c]{@{}c@{}}Accuracy 98.4\% \\ Precision 95.23\% \\ Recall 96.74\% \\ F1-score 95.98\% \\ Precision and recall \\ are low due to \\ complex model.\end{tabular} \\
\begin{tabular}[c]{@{}c@{}}\cite{Bangui2022}\end{tabular} &
  \begin{tabular}[c]{@{}c@{}}Hybrid model \\ based on RF\end{tabular} &
  \begin{tabular}[c]{@{}c@{}}KDD99, \\ CICIDS2017\end{tabular} &
  \begin{tabular}[c]{@{}c@{}}DoS, DDoS, \\ HeartBleed Etc.\end{tabular} &
  - &
  Accuracy 96.93\% \\
  \\
\begin{tabular}[c]{@{}c@{}}\cite{Polat2020}\end{tabular} &
  Softmax &
  \begin{tabular}[c]{@{}c@{}}Generate\\ (17779)\end{tabular} &
  DDoS &
  - &
  Accuracy 96.9\% \\
\begin{tabular}[c]{@{}c@{}}\cite{Alsarhan2021}\end{tabular} &
  \begin{tabular}[c]{@{}c@{}}SVM, GA, \\ PSO, ACO\end{tabular} &
  NSL-KDD &
  \begin{tabular}[c]{@{}c@{}}DoS, Probing, \\ Unauthorized \\ access\end{tabular} &
  - &
  Accuracy 98\% \\
\begin{tabular}[c]{@{}c@{}}\cite{Gao2019}\end{tabular} &
  RF &
  \begin{tabular}[c]{@{}c@{}}NSL-KDD, \\ UNSW-NB15\end{tabular} &
  DDoS &
  - &
  Accuracy 99.95\% \\
\begin{tabular}[c]{@{}c@{}}\cite{Alladi2021}\end{tabular} &
  \begin{tabular}[c]{@{}c@{}}CNN-LSTM, \\ CNN- bidirectional \\ LSTM, LSTM, \\ Auto-encoders \\ Stacked LSTM, \\ GRU\end{tabular} &
  VeReMi &
  \begin{tabular}[c]{@{}c@{}}Various types \\ of Anomalies\end{tabular} &
  - &
  Accuracy 98\% \\
Proposed Framework &
  \begin{tabular}[c]{@{}c@{}}RF, CatBoost,\\ LR\end{tabular} &
  VeReMi &
  \begin{tabular}[c]{@{}c@{}}Various attacks \\ (Constant, Random,\\ Eventual Stop)\end{tabular} &
  \begin{tabular}[c]{@{}c@{}}Stacking \\ Ensemble with \\ SHAP Analysis\end{tabular} &
  \begin{tabular}[c]{@{}c@{}}Accuracy: 99.68\%, \\ Precision: 99.6\%, \\ Recall: 99.31\%, \\ F1-Score: 99.45\%\end{tabular} \\
\\  \hline
\end{tabular}
\end{table*} 

In their work, \cite{Alheeti2017} using Linear and Quadratic Discriminant Analysis, the authors present an intelligent IDS to guard against assaults, namely DoS and Black Hole attacks. To begin, evil actions are modelled and then emulated. Next, the authors make several adjustments to the AODV protocol to produce the DoS attack. In this scenario, denial-of-service attacks are carried out by dropping packets. After that, multiple mobilities and traffic setups were developed. SUMO and MOVE produced a realistic setting of both abnormal and typical behaviour. In addition to that, ns-2 was utilised to imitate various forms of communication. The dataset is derived from the ns-2 trace file, and all 21 characteristics are preserved throughout the process. The data is given a fuzzy appearance before being put through the test and the training. The findings of the Linear Discriminant Analysis indicate a detection rate of 86.44\%, whereas the Quadratic Discriminant Analysis results indicate a detection rate of 85.67\%. 

In the study conducted by \cite{Sharma2018}, the authors suggest a novel technique for selecting a steady Cluster Head, and it uses Hybrid Fuzzy Multi-standards. After that, an ML-based intrusion detection system that uses SVM is utilised to identify malicious activities. The utilisation of a Dolphin Swarm Algorithm results in an improvement to the SVM-based IDS detection capabilities. This program uses dolphin swarms' hunting and preying behaviour to locate and isolate misbehaving nodes in a network. The future method is evaluated through the use of simulation. Matlab and NetSim simulate the network, while SUMO models the traffic. During the simulation, a range of node densities from 50 to 300 are examined, with a maximum of 45\% of the cars acting as attackers.~The suggested intrusion detection system has a detection rate greater than 98\%, selective forwarding, and wormhole attack; however, this percentage can vary based on the number of vehicles.

As outlined by \cite{Kumar2015}, a cloud intrusion detection system founded on clustering, and Learning Automate (LA) was presented. This method can potentially improve detection activity. The LA determines the leadership of clusters based on factors such as the total comparative velocity and the connectedness grade. The next stage, safeguarding the data, will begin after this procedure. In this process stage, an automaton will validate messages using the HMAC algorithm. When utilising the method that has been suggested, around 93\% of harmful actions are recognised, and when mobility is added to leadership formation, the percentage of false positives is reduced. In addition to that, this technique is flexible enough to adapt to any alterations made to the network's nodes. 

The work of \cite{Shu2021} presented an IDS for VANETs by implementing a Software-Defined Networking (SDN)-Based controller on each base station to differentiate between regular network traffic and attack network flows. Using the total network flow data, they trained several SDN controllers for the full VANET using GAN. This IDS approach permits distributed SDN controllers to detect their sub-network flows directly, and it can reduce high processing overheads. The author tested their algorithm using criteria along with precision, Accuracy, F1-Score, recall, and area under the curve (AUC), where maximum accuracy is 98.4\%.

As described by \cite{Zeng2019}, a deep learning-based intrusion detection method incorporating CNN and LSTM. The proposed model's efficiency was compared to previously developed ML methods using a public dataset and an NS3 simulated VANET dataset with 28236 entries. The recommended model earned an F1 score ranging from 91\% to 99\%.

Before analysing the above literature, it is essential to consider the findings reported by the National Institute of Standards and Technology (NIST) in 2007: history proved that high rates of false positives and negatives are allied with network-based IDPSs. NIDPSs cannot identify threats inside encoded network traffic; consequently, the system must be installed where the system can screen traffic before encoding or after decoding, or host-based IDPSs must be employed on endpoints to monitor unencrypted movement. Furthermore, NIDPSs are frequently incapable of performing complete analysis under high loads. Furthermore, it requires enough tuning to get good detection accuracy. \cite{Scarfone2007}. In 2022, NIDSs improved to analyse under full load when passing high traffic flow due to modern hardware specs and ML algorithms being used effectively. Besides, researchers should conduct their experiments with an extensive database to get at least 100\% accuracy for known attacks. The literature review notes two types of IDS research: A) The ML-based model is built with performance analysis. And B) Innovated Model implementation in a simulator to make a VANET environment to check feasibility before real-time practice.   

In their analysis, \cite{Gad2021} discussed ML-based IDS techniques in VANET that are impressive. Researchers used SMOTE, chi2, and SMOTE+chi2 to eliminate the database imbalance problem in the feature engineering phase, and the XGBoost method's overall score is better than others. In the Training-Testing phase, researchers used Logistic regression (LR), naive Bayes (NB), decision tree (DT), support vector machine (SVM), k-nearest neighbour (kNN), Random Forest (RF), AdaBoost, and XGBoost. This experiment could be better by using the Ensemble Sacking classification to increase overall performance, which this paper wants to address.  

In the research conducted by \cite{Türkoğlu2022}, the Minimum Redundancy Maximum Relevance (MRMR) feature selection algorithm to get the most correlated features for attack detection. First, scholars split the dataset in 3 ways: Training, Validation, and Test Data. After that, researchers used Bayesian Hyperparameter Optimization with training and validation data. However, 3-way splitting (60\%, 20\% and 20\%) was not mandatory for tuning. Instead, the same training data could be used for tuning purposes. That way, 70\% or 80\% of data could be used for training the database, and 30\% or 20\% of data will be used for testing purposes to increase the overall accuracy.

As explored by \cite{Aneja2018}, their research uses ANN (feed-forward network) with backpropagation technique. This backpropagation operates by scheming the gradient at every layer, known as the local gradient. Then, the weights and bias values are updated according to Levenberg-Marquardt optimisation. Moreover, a Genetic algorithm is used for feature selection.

In the study conducted by \cite{Alsarhan2021}, the researchers introduce an optimised Support Vector Machine (SVM)-based Intrusion Detection System (IDS) for Vehicular Ad Hoc Networks (VANETs), employing Genetic Algorithm (GA), Particle Swarm Optimisation (PSO) and Ant Colony Optimization (ACO) to enhance accuracy. GA outperformed other algorithms, demonstrating improved detection rates and minimised false positives. A comprehensive evaluation validates the approach on the NSL-KDD dataset, providing significant insights into VANET security. Table \ref{table:review} provides a summary of the key works and findings in relation to ML-based IDS models in VANET and SDVN.

\begin{figure}[t]
    \centering
    \includegraphics[width=0.35\textwidth]{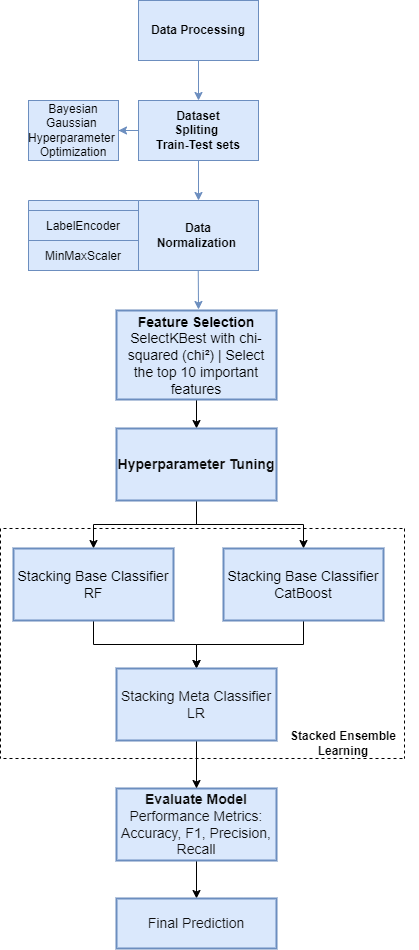} 
    \caption{Binary Classification.}
    \label{fig: Binary Classification}
\end{figure}

\begin{figure}[t]
    \centering
    \includegraphics[width=0.35\textwidth]{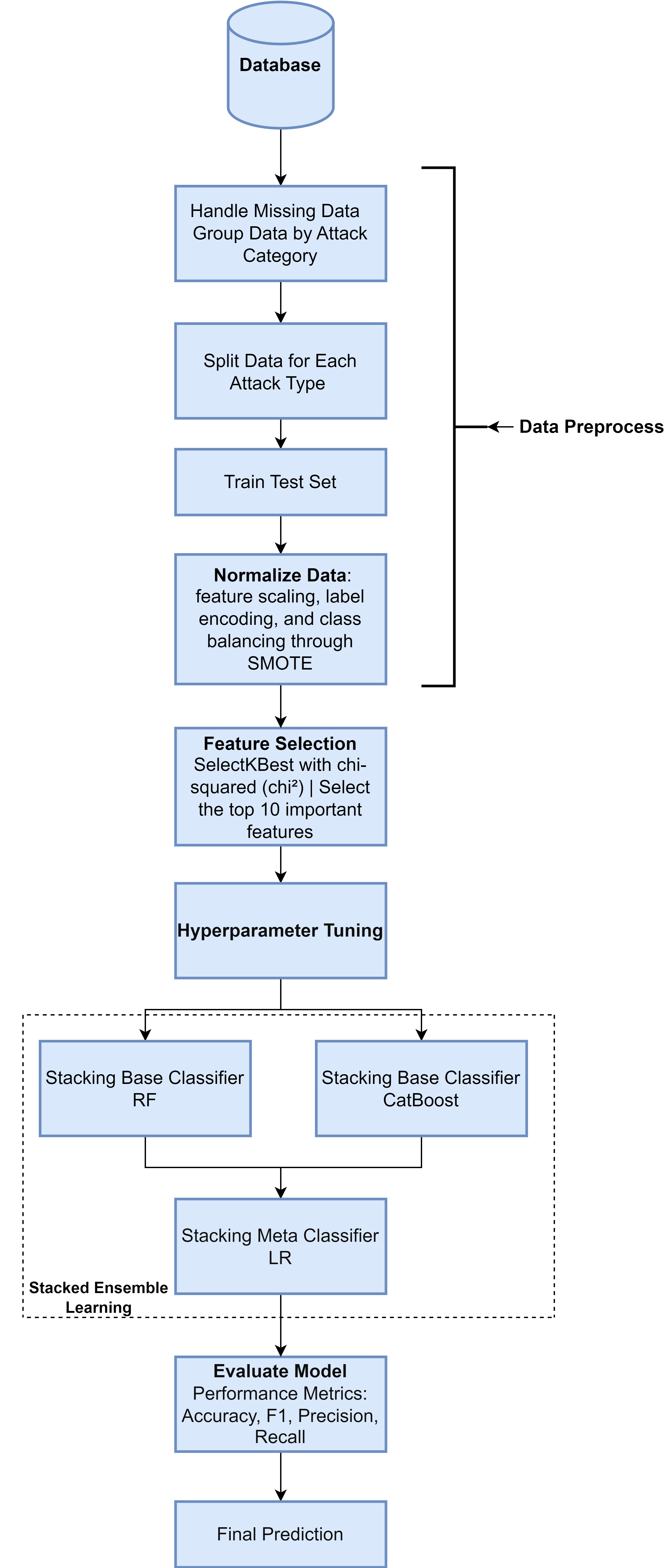} 
    \caption{Multiclass Classification (1-Vs-1).}
    \label{fig: multiclass-classification}
\end{figure}

\subsection{Advancing SD-VANETs with Explainable AI}

Recent advancements in IDS tailored for SD-VANETs have leveraged a range of machine learning (ML) models. However, few of these systems have incorporated XAI to clarify their predictions. Prominent approaches, such as CNNs and RNNs, have been employed for anomaly detection (\cite{Zeng2019-yr}). However, these models frequently lack the transparency required for use in safety-critical environments, which can significantly diminish their effectiveness.

Emerging research has underscored the potential of XAI methodologies, particularly federated learning. This approach facilitates collaborative model training while ensuring data privacy and interoperability (\cite{Huang2023-my}). By utilizing decentralized data, federated learning not only boosts model accuracy but also sheds light on individual model contributions and decision-making processes, making it particularly relevant for the cybersecurity challenges faced by SD-VANETs.

In response to these challenges, our study seeks to enhance model interpretability by integrating SHAP analysis. This method clarifies feature importance and improves the transparency of model predictions. Additionally, we tackle persistent issues in IDS, such as elevated false positive rates and the identification of unknown attacks, by employing a stacked ensemble learning strategy. This approach has proven effective in enhancing overall performance through increased feature diversity and optimized decision boundaries, as evidenced in recent research by (\cite{Wang2024-yo} and \cite{Kim2022-ti}).

By merging SHAP analysis with ensemble techniques and leveraging the advantages of federated learning, our framework aims to deliver robust security solutions for SD-VANETs while promoting trust and accountability in automated decision-making processes.

\section{Technical Implementation}\label{sec4}
The VeReMi dataset was created to test the effectiveness of VANET misbehaviour detection systems (vehicular networks). This database contains onboard unit message logs created from a simulation environment and tagged ground truth. The collection contains malicious messages that are meant to cause erroneous application behaviour, which misbehaviour detection techniques are designed to stop. The initial dataset includes five types of position falsification attacks. This data was generated by \cite{Van_der_Heijden2018-ah}, and It has been collected from \cite{Sharma2021}.

To achieve the best possible hardware support for running the code, Microsoft Azure ML Studio Virtual Machine (VM) instance Standard E4s v3 (4 cores, 32 GB RAM, 64 GB HDD) and Processing unit CPU - Memory optimised were utilised for this experiment. 

\subsection{Supervised ML Approaches }

Specifically, the RF classifier, LR, and CatBoost have been examined among the ML systems in this study. 

\paragraph{Random Forest Classifier:} The Random Forest Classifier is an effective machine learning algorithm that is part of the ensemble learning category. It works by creating numerous decision trees during the training process and combining their predictions to enhance overall accuracy. Each tree is built using a random sample of the data, which helps to introduce variety and minimize the chances of overfitting. The final classification result is determined by a majority vote, where the most frequently predicted class among all the trees is chosen. This method not only improves accuracy but also adds resilience against noise and fluctuations in the dataset. Random Forest classifiers are popular across various fields, such as finance, healthcare, and cybersecurity, because they can manage large datasets with many features and provide insights into the importance of different features, making them both interpretable and effective for tasks involving classification and regression.

\paragraph{Logistic Regression:}
\noindent LR is a linear model used to classify instead of predict. A logistic function is used for probabilities that describe the possible results of a single trial in this model. Typically, a sigmoid curve with the shape of a Sigmoid curve is called a logistic function or logistic curve (\cite{Wikipedia2023}).

\paragraph{CatBoost:} As noted by \cite{dorogush2018}, CatBoost is a gradient-boosting algorithm developed by Yandex that excels in handling categorical features, making it particularly suitable for datasets with heterogeneous data types. Unlike traditional boosting methods, CatBoost employs an innovative approach called ordered boosting, which mitigates overfitting by utilizing a permutation-based method for training. Its efficiency is further enhanced through support for both numerical and categorical data, eliminating the need for extensive preprocessing. With state-of-the-art performance across various machine learning tasks, CatBoost has gained popularity in both academia and industry, offering robustness and interpretability in model training and evaluation, particularly in the context of complex datasets.

In essence, Random Forest is particularly strong at managing high-dimensional data and is known for its resilience against overfitting. On the other hand, CatBoost shines when dealing with categorical features and effectively reduces gradient bias. By stacking these two models, we aim to improve both detection performance and interpretability within VANETs.

\section{Methodology}

\subsection{Technical Implementation Methodology}
This study addresses multiclass classification problems using two approaches: binary decision reduction and multiclass methodology. The performances of these two approaches will be discussed below.

\subsection{Data Processing}

\subsubsection{Binary Decision Reduction}
Before inputting the data into machine learning algorithms, it is essential to perform cleaning and preparation steps to ensure high performance. Unnecessary features may detract from the chosen ML method's effectiveness. We handled missing values, which are common in datasets, by replacing NaNs and infinities with appropriate substitutes. After grouping similar attacks and splitting the dataset into training and testing parts, we employed a binary classifier to label the dependent columns as either BENIGN or ATTACK. LabelEncoder normalized the labels, ensuring that the dataset contained values only between 0 and \(n_{\text{classes}}-1\). Additionally, MinMax Scaling was applied to mitigate the effects of variables measured on different scales, helping to prevent bias.

\subsubsection{General Multiclass Methodology}
Distinct datasets were constructed by isolating specific attacks while discarding the remainder, thereby creating unique data combinations for each attack versus BENIGN data.

\subsection{Feature Selection}
The SelectKBest class from the scikit-learn library employs various statistical tests to identify specific features. In this experiment, the chi-squared (\(\chi^2\)) statistical test was utilized to select the ten best features based on non-negative characteristics. The Chi-Square test quantifies the disparity between observed and expected data, allowing us to assess whether the discrepancies between categorical variables result from chance or an underlying association.

\subsection{Hyperparameter Optimization: Bayesian Optimization with Gaussian Process}
BayesSearchCV integrates ``fit'' and ``scoring'' methods, optimizing parameters through a cross-validated search of the parameter settings. Unlike GridSearchCV, which tests all parameter values, this approach experiments with several parameter settings drawn from selected distributions. The number of iterations (\(n_{\text{iter}}\)) determines how many parameter settings will be assessed, leading to a more efficient exploration of the hyperparameter space.

The hyperparameters can be continuous, discrete, or categorical. In this study, we implemented hyperparameter optimization for the stacking base classifier.

\subsection{Ensemble Stacking Classifier}
Ensemble learning encompasses three primary methods: bagging, boosting, and stacking. Bagging and boosting utilize alternative voting methods, while stacking serves as a versatile framework, integrating multiple ensemble approaches. This study utilized a stacking approach with Random Forest and CatBoost as base learners and Logistic Regression as the final estimator. The configuration for the stacking classifier is as Table \ref{tab:estimators_multiclass}:

\begin{table*}[tbp]
\fontsize{9}{9}\selectfont
\caption{Estimators for Multiclass Approach.}
\label{tab:estimators_multiclass}
\begin{tabular}{ll}
\hline
\textbf{Datasets} & \textbf{Multiclass Classification} \\ 
\hline
VeReMi & 
  \begin{tabular}[c]{@{}l@{}} 
  \texttt{estimators = \{('rf', RandomForestClassifier(}\\[0.2em]
        \texttt{n\_estimators=50, max\_depth=20,}\\[0.2em]
        \texttt{min\_samples\_split=10, min\_samples\_leaf=2,}\\[0.2em]
        \texttt{criterion='entropy', n\_jobs=-1)),}\\[0.2em]
        \texttt{('catboost', CatBoostClassifier(}\\[0.2em]
        \texttt{iterations=100, learning\_rate=0.1,}\\[0.2em]
        \texttt{depth=6, verbose=0)\}} \\[0.5em]

  \hline
  \texttt{clf = StackingClassifier(}\\[0.2em]
        \texttt{cv=3, estimators=estimators,}\\[0.2em]
        \texttt{final\_estimator=LogisticRegression(}\\[0.2em]
        \texttt{solver='lbfgs', max\_iter=1000),}\\[0.2em]
        \texttt{stack\_method='auto', passthrough=False, n\_jobs=-1)} 
  \end{tabular} \\ 
  \hline
\end{tabular}
\end{table*}

The choice of base learners is crucial for the stacking process, highlighting the importance of employing multiple models rather than relying on a single one. The stacking method uniquely employs meta-level learning for the final decision-making process.

\subsection{Experimental Setup}
With the help of the VeReMi datasets, our intrusion detection system employs ensemble learning classifiers to tackle multiclass classification challenges. 

The VeReMi dataset comprises multiple CSV files capturing different types of attacks. Data preprocessing involved eliminating records with null values (NaNs or Infinities) to ensure model reliability. Labels were transformed from numerical codes to meaningful descriptions (e.g., ``Constant Attack'', ``Random Attack''). The dataset was divided into training and validation sets in an 80:20 ratio to ensure a balanced representation of attack types. Normalization using MinMaxScaler ensured that all features were scaled appropriately, thereby eliminating visual biases in the model's predictions. Our approach involves combining models like RandomForest and CatBoost, with Logistic Regression serving as the meta-classifier. To address underrepresented attacks, cross-validation folds were set to CV = 3, helping to prevent overfitting against smaller sample sizes.

\section{Results and Evaluation}\label{sec5}

In our proposed model, accuracy serves as the primary metric for assessing performance within the Intrusion Detection System (IDS). The evaluation metrics include accuracy (ACC), detection rate (DR), precision, and recall, based on true positive (TP), true negative (TN), false positive (FP), and false negative (FN) counts.

The performance evaluation leverages distinct datasets for the training and testing phases of anomaly detection. As evidenced by the results, the achieved accuracy percentage is notably high, warranting recognition as outstanding. Additionally, the other performance metrics exhibit similarly encouraging results.

\begin{table*}[tbp]
\footnotesize
\caption{Performance evaluation for Binary Classification}
\label{tab:performance}
\begin{tabular}{llllll}
\hline
Datasets & Accuracy & Precision & Recall & F1 \\ \hline 
VeReMi & $99.68\%$ & $99.60\%$ & $99.31\%$ & $99.45\%$ \\ \hline 
\end{tabular}
\end{table*}

\begin{table*}[tbp]
\footnotesize
\caption{Model performance evaluation using Multiclass Approach for VeReMi dataset.}
\label{tab:multiclassperformance}
\begin{tabular}{llllll}
\hline
Attacks & Accuracy & Precision (\%) & Recall (\%) & F1 (\%) \\ \hline
BENIGN & $97.65\%$ & $100\%$ & $98\%$ & $99\%$ \\
Constant Attack & $100\%$ & $100\%$ & $100\%$ & $100\%$ \\
Constant Offset Attack & $99.74\%$ & $99\%$ & $100\%$ & $99\%$ \\
Eventual Stop Attack & $97.14\%$ & $78\%$ & $97\%$ & $87\%$ \\
Random Attack & $99.95\%$ & $100\%$ & $100\%$ & $100\%$ \\
Random Offset Attack & $99.10\%$ & $98\%$ & $99\%$ & $99\%$ \\ \hline
Accuracy & $98.11\%$ & & & \\
Macro Avg & $97.64\%$ & $96.17\%$ & $99.57\%$ & $97.14\%$ \\
Weighted Avg & $98.11\%$ & $98.29\%$ & $98.38\%$ & $98.28\%$ \\ \hline
\end{tabular}
\end{table*}

\subsection{Experimental Results}

The evaluation of the VeReMi dataset provides comprehensive insights into the performance of both binary and multiclass classification models. The binary classification approach achieved remarkable metrics, with accuracy reaching $99.68\%$, precision at $99.60\%$, recall at $99.31\%$, and an F1 score of $99.45\%$ for the VeReMi dataset. In contrast, the multiclass classification strategy adeptly identified various attack types, with the model achieving an overall accuracy of $98.11\%$. The lowest accuracy observed was $97.14\%$ for the Eventual Stop Attack, highlighting an area for potential improvement.

False negatives (FN) and false positives (FP) are critical indicators of a classification model’s success, as they represent the model's ability to accurately identify both positive and negative instances. An effective model should maintain low FN and FP rates, reflecting its capacity to correctly detect positive and negative cases. Our model demonstrated exceptional performance, particularly in multiclass classifications. The estimator table outlines the parameters applied to achieve these impressive results.

\begin{figure*}[tbp] \centering \includegraphics[width=0.6\textwidth]{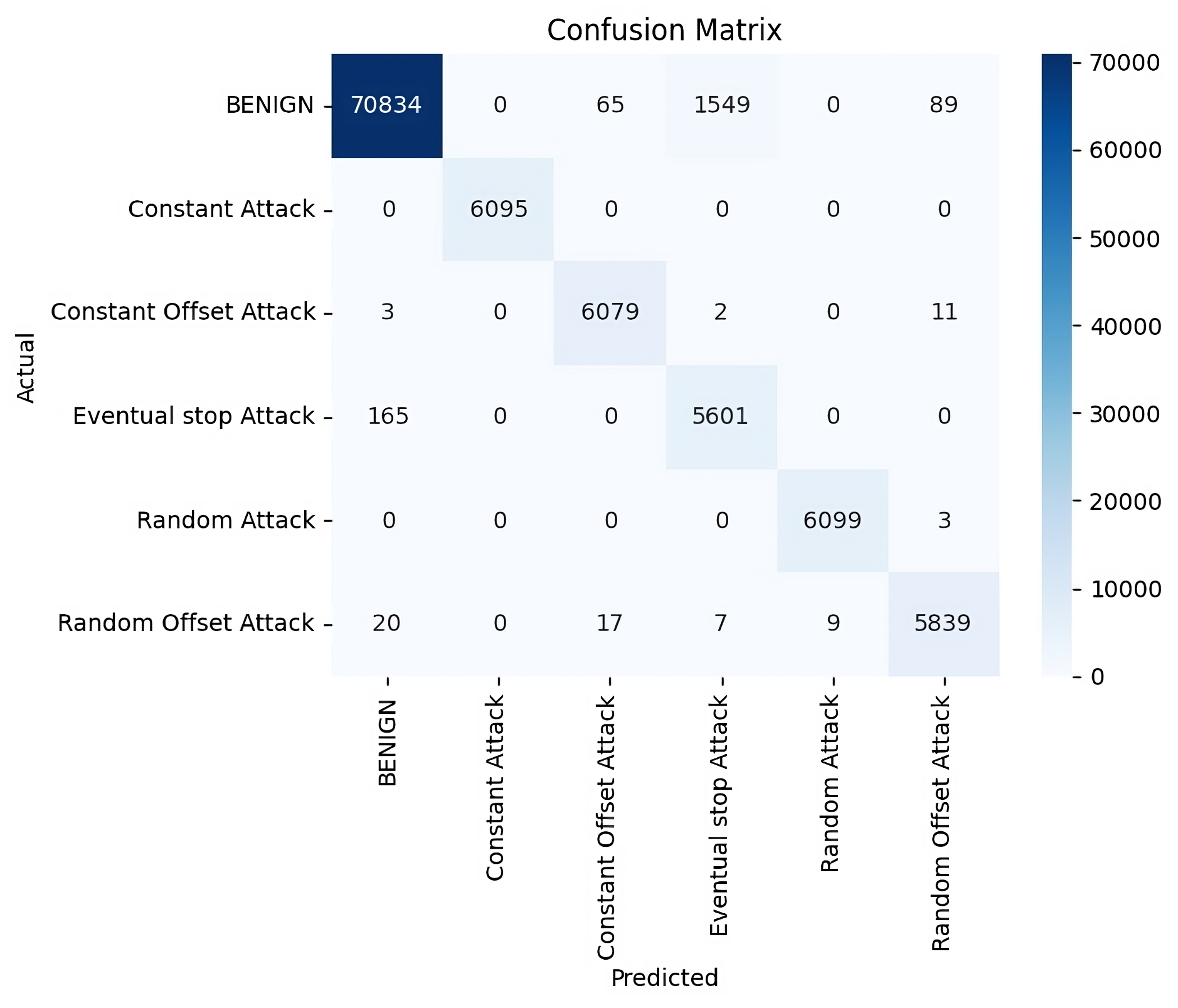} \caption{Confusion Matrix by Attacks for Multiclass Classification.} \label{fig:3} \end{figure*}

The slight performance variances observed in attack detection within the IDS model using the VeReMi dataset can be further elucidated through metrics such as precision, recall, F1 scores, false positives, and false negatives. As indicated in the dataset, the model exhibited slightly lower precision and higher false negatives for the Eventual Stop Attack, suggesting inherent challenges in consistently detecting this specific attack pattern. This difficulty may stem from the less straightforward characteristics associated with the Eventual Stop Attack. Conversely, the high precision scores for the Random and Constant Offset attacks suggest that these attack patterns present consistent features, enabling the model to learn them efficiently.

Moreover, false negative rates indicate a balance between specificity and sensitivity within the model. The elevated false negative rate for the Eventual Stop Attack implies that while the model may be precise in its predictions, it lacks the sensitivity necessary to detect this attack consistently. The use of Random Forest and CatBoost as base learners, complemented by Logistic Regression as a final estimator, has enhanced model effectiveness, particularly in evaluating performance against the VeReMi dataset. The strategic implementation of these algorithms allows for the leveraging of their strengths while mitigating their weaknesses through a stacking approach. The variance in model effectiveness across different attack types is influenced by factors such as the complexity and frequency of attack patterns, the degree of similarity among harmful and benign features, as well as the volume and quality of data utilized for training. Collectively, these factors play a crucial role in determining how effectively the model learns, ultimately shaping its accuracy in detecting and classifying various attack types.

In conclusion, both binary and multiclass classification models performed admirably in detecting attacks within the VeReMi dataset. The high accuracy, precision, recall, and F1 scores, alongside low RMSE and high $R^2$ values, affirm that the model's predictions were accurate and well-fitted to the data. The training duration varied according to the dataset size and the number of attack types classified.

To substantiate the robustness and efficiency of our approach, we conducted comparative analyses against other established methods, detailed in Table 3. For instance, \cite{Gad2021} achieved an accuracy of $99.1\%$ using XGBoost after feature selection; \cite{Türkoğlu2022} attained $99.35\%$ accuracy with a Decision Tree following MRMR feature selection; \cite{Zeng2019} realized $98.5\%$ accuracy with a combination of CNN and LSTM in DeepVCM; and \cite{Aneja2018} developed an ANN with a genetic algorithm, achieving $99\%$ accuracy. In contrast, our proposed stacked ensemble model surpassed these benchmarks, achieving an impressive accuracy of $99.68\%$ with the VeReMi dataset.

\begin{figure*}[tbp] \centering \subfigure[SHAP Analysis Plot for CatBoost]{\includegraphics[width=0.45\textwidth]{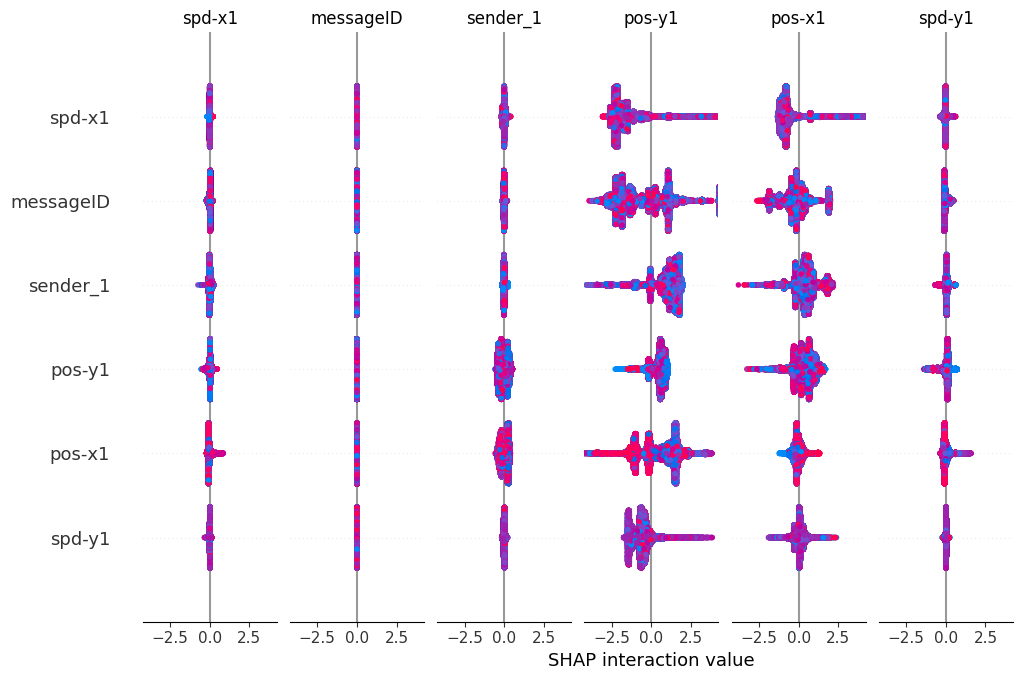}} \hfill \subfigure[SHAP Analysis Plot for Random Forest]{\includegraphics[width=0.45\textwidth]{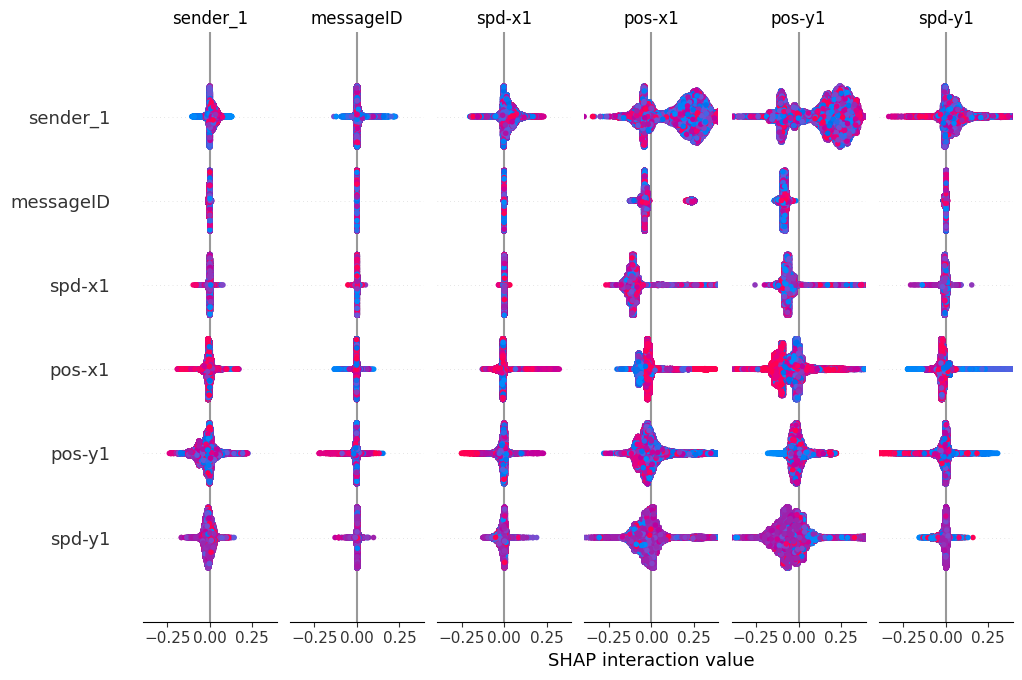}} \caption{Feature Impact and Misclassification Analysis for Multiclass Classification.} \label{fig
} \end{figure*}

\subsection{Feature Impact and Misclassification}
The confusion matrix reveals the model's strong capability in classifying network behaviours, particularly in accurately identifying benign traffic, with 70,834 correct classifications. However, there are instances where benign traffic is misclassified as Eventual Stop Attacks (1,549 cases), highlighting areas that need enhancement. On the other hand, categories such as Constant Attack, Constant Offset Attack, and Random Attack demonstrate remarkable accuracy, with very few misclassifications. Although the model correctly identifies the Eventual Stop Attack 5,601 times, it occasionally misclassifies this category as benign, signalling the necessity for improved detection strategies.

This experiment used SHAP analysis to assess how well the CatBoostClassifier and RandomForestClassifier models could distinguish between benign and attack traffic. The SHAP summary plots revealed important insights, showing which features impacted the model’s decisions most. For CatBoost, features like pos-y1, pos-x1, and spd-y1 stood out, especially when their values suggested abnormal activity. Similarly, RandomForest heavily relied on features like sender\_1, messageID, and spd-x1, where overlapping patterns between benign and attack data pointed to possible misclassification issues.

SHAP interaction plots allowed for further examination of feature independence and vulnerability. A notable interaction was found between pos-y1 and spd-x1, suggesting a dependency between these features. This raises a critical issue, as altering one feature could affect the other, leading to incorrect classifications. This is an important consideration, as attackers could potentially exploit these dependencies to deceive the model.

Additionally, the overlapping SHAP values between benign and attack data clarified why the model sometimes misclassified attack traffic as benign. These similarities in feature distributions between the two classes complicated the classification process. This suggests that using a two-step classification system—first identifying traffic as either benign or attack, and then classifying specific attack types—could help reduce the risk of critical errors, such as attacks being mistakenly identified as benign, which is a significant concern in cybersecurity.

The SHAP analyses also helped explain the misclassifications, as the similar feature importance between the classes indicated that the model’s errors were understandable. This highlights the need for improved feature selection or refining the decision boundaries, particularly for features where overlap occurs, to enhance the model’s ability to accurately detect attacks.

\section{Limitations and Future Scope}

\subsection{Expanded Limitations}

One of the major hurdles we encountered in this study stems from our reliance on SHAP analysis for understanding how our model makes decisions. While SHAP is a powerful tool for identifying the significance of different features, it doesn’t fully capture the complex relationships and dependencies that exist among them. For example, features like position coordinates (pos-y1) and speed (spd-x1) may show overlapping SHAP values, which can lead to confusion in classification tasks. This overlap can create ambiguity that misguides the model in critical situations, rendering it vulnerable to adversarial attacks that exploit these hidden connections.

\subsection{Future Directions}
As we look ahead, we see tremendous potential for our framework to evolve by incorporating encrypted data or data obfuscation. This move would not only enhance privacy but also ensure that threat detection remains robust. Furthermore, putting our model to the test on real-time data from VANETs could provide invaluable insights into its scalability and reliability in high-stakes, real-world scenarios. We also believe it is essential to explore alternative methods of explainability that prioritize feature independence. By addressing some of the shortcomings associated with SHAP analysis, we could better equip our model to handle more complex challenges. Ultimately, our goal is to build a more resilient security framework that meets the demands of intelligent transportation systems, paving the way for safer and smarter roadways.

\section{Conclusion}

We have developed a robust Intrusion Detection System (IDS) framework specifically designed for Software-Defined Networks (SDN)-based Vehicle Ad-hoc Networks (VANETs). What makes this research stand out is our innovative use of stacked ensemble learning combined with SHAP analysis, which not only enhances detection accuracy but also significantly improves interpretability. While many existing models tend to focus on either performance or explainability, our approach successfully balances both, addressing a critical need in the field. By leveraging Random Forest and CatBoost alongside Logistic Regression, our model achieved an impressive accuracy 99.68\% for binary classification and overall accuracy of 98.11\% for multiclass classification on the VeReMi dataset. The integration of SHAP analysis adds an essential layer of transparency, revealing how different features interact and helping to pinpoint vulnerabilities that previous models often miss. 

\section*{Declarations}
\subsubsection*{Competing interests}
The authors declare that they have no competing interests.
\subsubsection*{Authors' contributions}
Not applicable.

\subsubsection*{Availability of data and materials}
Not applicable. 
\subsubsection*{Funding}
Not applicable. 
\subsubsection*{Acknowledgements}
Not applicable. 
\medskip

\bibliographystyle{cas-model2-names}

\bibliography{article-v2}

\end{document}